\documentclass[preprint]{aastex}

\usepackage{natbib}
\usepackage{amsmath} \usepackage{amssymb}
\usepackage[colorlinks=true,urlcolor=black,linkcolor=black,citecolor=black]{hyperref}
\usepackage{ulem}

\shorttitle{Fast and Reliable Time Delay Estimation of Strong Lens Systems Using Method of Smoothing and Cross-Correlation}
\shortauthors{Aghamousa, Shafieloo}

 \usepackage{xcolor}
 \newcounter{attnctr} 

\begin{document}

\title{Fast and Reliable Time Delay Estimation of Strong Lens Systems Using Method of Smoothing and Cross-Correlation}

\author{Amir Aghamousa}
\affil{Asia Pacific Center for Theoretical Physics, Pohang, Gyeongbuk 790-784, Korea}
\email{amir@apctp.org}

\author{Arman Shafieloo}
\affil{Korea Astronomy and Space Science Institute, Daejeon, 305-348 Korea}
\email{shafieloo@kasi.re.kr}

\begin{abstract}
The observable time delays between the multiple images of strong
lensing systems with time variable sources can provide us with some
valuable information to probe the expansion history of the Universe.
Estimation of these time delays can be very challenging due to
complexities of the observed data where there are seasonal gaps, various
noises and systematics such as unknown microlensing effects. In this
paper we introduce a novel approach to estimate the time delays for
strong lensing systems implementing various statistical methods of
data analysis including the method of smoothing and cross-correlation. The
method we introduce in this paper has been recently used in TDC0 and
TDC1 Strong Lens Time Delay Challenges and has shown its power in
reliable and precise estimation of time delays dealing with data with
different complexities.
\end{abstract}

\keywords{}

\section{Introduction}\label{introduction}
Over the last few decades cosmology has emerged as a precision science where we have access to a variety of data from different resources. Multiple images of strong lensing systems have shown to be a rich source of information to extract some precise knowledge about the expansion history of the Universe and break the degeneracies between different cosmological models \citep{Eric2004,Suyu2013}. 

In strong gravitational lensing not only we do have multiple images of the same source, but since the path of light is different for different images there will be a time delay between these images. When the source is variable we might be able to estimate the time delays between different images. If we can also obtain proper information about the strong gravitational lens system and model it appropriately (to estimate the gravitational potentials different images have passed through), then we can use all these information to measure cosmic distances between the source, strong gravitational lens system and us and this can be used to reconstruct the expansion history of the Universe \citep{Refsdal1964, Kochanek2002}. There have been recent advances in both fronts: in modeling of strong lens systems\citep{Oguri2007, Fadely2010, Suyu2009, Suyu2010} and also there have been programs in long term observations of the multiple images of some strong lensing systems in order to estimate time delays between the multiple images \citep{Courbin2011, Eulaers2013, Tewes2013b, Kumar2013}. The importance of analyzing strong lens systems in order to reconstruct the expansion history of the Universe and their broad cosmological implications is becoming increasingly evident for the cosmological community. This is due to the fact that one can use these systems to measure the expansion history of the Universe in a completely different way to standardized candles and rulers such as supernovae type Ia or baryon acoustic oscillations. Hence strong lensing systems can be used as a complementary approach to study the expansion of the Universe and even to calibrate the other probes of the expansion history \citep{Eric2004, Eric2011, Suyu2012}.

From the observational point of view, the COSMOGRAIL project (http://www.cosmograil.org) is going to yield tens of time delays for strong lens systems and in the future the Large Synoptic Survey Telescope (LSST) \citep{LSST2009, Ivezic2008} will be monitoring several thousand time delay lens systems \citep{LSST2012, Oguri2010}. This assures that there will be enough information in the near future to do precision cosmology using strong lensing systems, if we can model the lens systems appropriately and estimate the time delays precisely. From the theoretical point of view, while there have been important advances in modeling of strong lens systems, there have been also various attempts to provide efficient and robust algorithms in order to estimate the time delays between strong lens system images with high precision and accuracy \citep{Tewes2013a,Tewes2013b,Suyu2013,Hojjati2013}. These proposed algorithms are all very different in nature, and they all have both advantages and some weaknesses. In an effort to move towards more accurate and precise algorithms of time delay estimation, a group of scientists organized the Time Delay Challenge (TDC) to encourage different scientific groups to modify or propose new algorithms for time delay estimation \citep{TDC0-2013,TDC1-2014}. In this challenge some simulated data were prepared at two stages, TDC0 and TDC1 and participants were requested to perform their own analysis and report the estimated time delays and corresponding uncertainties. As one of the participating groups we joined the challenge and developed our own fully automated and reliable algorithm to estimate time delays (for more details about TDC1 and proposed algorithms by participated groups see \cite{TDC1-2014}). 
Our algorithm is also fast in analyzing the light curves. The whole processing time of a pair of light curves by using a typical PC is of the order of few minutes. We use the \texttt{R} programing language \citep{R} in both computational and plotting tasks.
In this paper we introduce and explain the algorithm that we used in TDC1 challenge.    

In what follows, section~\ref{sec:data} describes TDC1 simulated data. In section~\ref{sec:methodology} we explain the methodology which contains the smoothing method and cross-correlation and the strategy we follow to estimate the time delays. We discuss our fast and reliable approach to estimate the uncertainties of the recovered time delays and finally in section~\ref{sec:results_discussion} we present the results and end with a discussion.

\section{Data}~\label{sec:data}

\begin{figure}
   \includegraphics[width=0.9\textwidth]{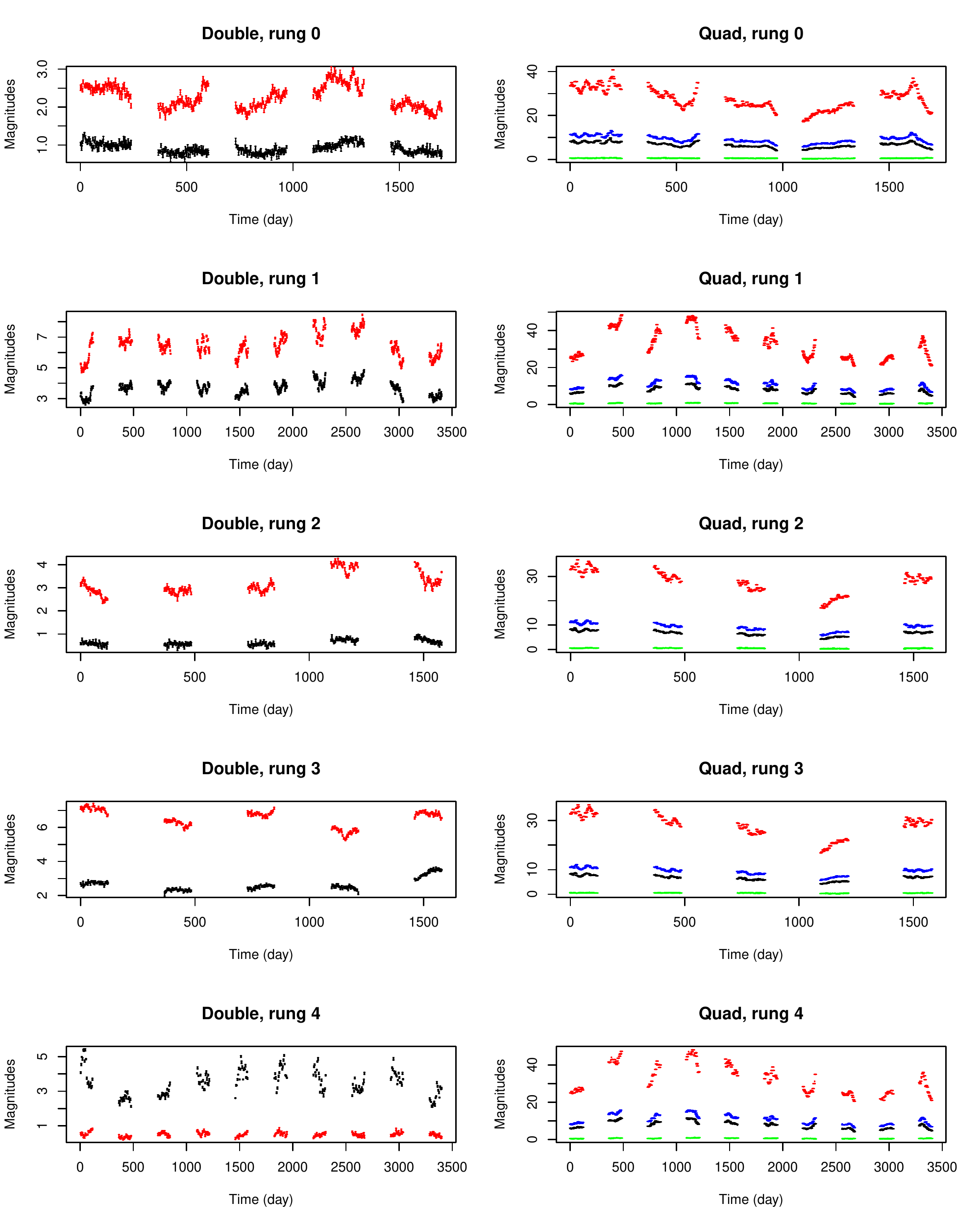}
  \caption{\label{fig:data} The light curves of a typical Double system (pair 1 for each rung) and a Quad system (pair 1A and pair 1B for each rung) from TDC1 simulated data is presented.}
 \end{figure}

The TDC1 simulated data is provided in five different categories (rungs). Each rung contains the light curves of 720 Double and 152 Quad image systems. Tables~\ref{tab:double_quad} lists the properties of the data in each rung. For each rung and within all data samples, the number of data segments (segments are separated where there is a gap of more than 100 days between two neighboring data points.), minimum and maximum number of data points, minimum and maximum length of data, minimum and maximum of the shortest and longest gaps between two data points are tabulated. We see the Doubles and Quads in each rung have nearly the same properties. Figure~\ref{fig:data} depicts typical light curves for each rung. From Figure~\ref{fig:data} and the tables we note that the data is not equi-spaced in time and also suffers from gaps in some intervals and in some cases there are evident microlensing effects. These specifications ensure that there is a need for sophisticated and perhaps complicated statistical approaches in order to correctly estimate the time delays. In section~\ref{sec:methodology} we elaborate the methodology that we employ to deal with this data.


\begin{table}[!htb]
\begin{center}
\vspace{6pt}
\resizebox{15cm}{!} {
\begin{tabular}{|l|l|l|l|l|l|l|}
\hline
System & Rung  &  Number of data segments &  Number of data points &  Length of data (day)   & Min. interval (day) &    Max. interval (day)\\
\hline
Double & 0   &     5   &   (377, 420)  &  (1691.85, 1699.89)  & (0,    0.82)   &    (126.72, 134.09) \\
Double & 1   &    10   &  (382, 418)   &  (3396.42, 3404.82)   & (0 ,   0.82)   &    (247.85, 255.19) \\
Double & 2   &     5   &   (199, 199)  &  (1578.00, 1578.00)     & (3,    3)      &    (249, 249)  \\
Double & 3   &     5   &  (185, 215)   &  (1570.89, 1579.80)  & (0,    1.24)   &    (246.68, 255.24) \\
Double & 4   &    10   &  (192, 212)   &  (3391.38, 3404.68)  & (1.70, 4.24)   &   (250.97, 261.17) \\
\hline
Quad & 0   &   5     & (382, 415)  &  (1691.47, 1699.77)  & (0  ,  0.76)   &   (126.47, 133.14)\\
Quad & 1   &   10    & (372, 414)  &  (3396.48, 3404.78)  & (0  ,  0.82)  &    (248.26, 254.23)\\
Quad & 2   &    5    & (199, 199)  &  (1578.00, 1578.00)  & (3 ,   3)     &    (249  ,  249)\\
Quad & 3   &    5    & (188, 216)  &  (1571.89, 1579.81)  & (0 ,   0.99)  &    (247.11, 253.14)\\
Quad & 4   &   10    & (192, 207)  &  (3391.62, 3403.89)  & (2.16 4.08)  &    (251.29, 260.12) \\
\hline
\end{tabular}
}
\end{center}\caption{~\label{tab:double_quad} The summary of TDC1 simulated data (Double and Quad systems). For each rung and within all data samples, number of data segments, minimum and maximum number of data points, minimum and maximum of length of data, minimum and maximum of the shortest and longest gaps between two neighboring data points are tabulated. Furthermore, the errors on the magnitudes of light curves are similar for all rungs of Double and Quad systems in the ranges of $\sim (0.001$ to $0.128)$ and $\sim (0.001$ to $0.135)$ respectively.}
\end{table}

\section{Methodology}\label{sec:methodology}

To find the time delay between two light curves we need somehow to make a comparison between curves. We employ an application of cross-correlation to find the lagged time between the two light curves. We argue that we should get the highest cross-correlation between the two light curves if one of them is shifted in time coordinate approximately the actual time delay between the curves. This seems trivial since all light curves are coming from the same source and they should have similar features in them. But as we described in the previous section the light curve data are not equi-spaced in time axis, with several seasonal gaps. This sparseness of the data can in fact cause inaccurate or even incorrect estimation of time delay. Therefore in practical cases one cannot use cross-correlation directly applied to the data to estimate the time delays in strong lens systems. To tackle this issue, before using cross-correlation for comparing the light curves in each system we obtain a proper smoothed forms of the light curves. Then we derive the cross-correlations between various light curves and the smoothed forms. 
In this section first we introduce the smoothing method and the cross-correlation that we use to estimate the time delays and in continuation we propose a fast and efficient algorithm for error estimation.

\subsection{Smoothing}\label{subsec:smoothing}
We adopt a model-independent smoothing method introduced by \cite{Arman2006,Arman2007,Arman2010} and \cite{Arman2012}. In this method the smooth light curve, $A^{s}(t)$, is obtained at any required time $t$ by an iterative procedure:
\begin{equation}
A^{s}(t)= A^{g}(t) + N(t)\sum_{i}\frac{A^{d}(t_i)-A^{g}(t_i)}{\sigma^{2}_{d}(t_i)} \times exp\left [ - \frac{(t_i - t)^2}{2\Delta^2} \right ]
\end{equation}
where
\begin{equation}
N(t)^{-1}= \sum_{i} exp \left [ - \frac{(t_i - t)^2}{2\Delta^2} \right ] \frac{1}{\sigma^{2}_{d}(t_i)}
\end{equation}
where $A^{d}(t_i)$ and $\sigma_{d}(t_i)$ represent the $i^{th}$ data point and its uncertainty, $A^{g}(t)$ is the guess model, $N(t)$ is the normalization factor and $\Delta$ is width of smoothing.
In this procedure the first iteration starts with an initial guess model, $A^{g0}(t)$, which results the smooth fit, $A^{s1}(t)$. In the next iteration the obtained smooth fit is used as the next guess model $A^{g1}(t)=A^{s1}(t)$ which yields to the new smooth fit $A^{s2}(t)$. The procedure can be continued to future iterations. It has been shown before that results are independent of the assumption of initial guess model \citep{Arman2007}. We should also note that the size of $\Delta$ represents the minimum size of the fine structure in the data that will be finally presented in the smoothed reconstruction. In this procedure, 
the smoothing width $\Delta$ has also a direct connection to the speed of convergence to the final required smooth fit. In fact in order to reduce the time of computation it is important to find the desirable combination of $\Delta$ and number of required iterations in the procedure. We select the optimum value of $\Delta$ and number of iterations by simulating data based on some of the TDC0 data. In our procedure we set $\Delta=8$ days and $3$ iterations. By choosing a small value of $\Delta$ we reduce computation time by having very low number of iterations in the procedure. Based on our simulations we noticed that choosing a smaller value of $\Delta$ result to noise fitting rather than finding actual features in the light curves and on the other hand choosing a much larger value of $\Delta$ could smooth out the light curves significantly and wash out the important features in the data. Our simulation results indicate that while $\Delta=8$ seems to be a reasonable value, in the case of the TDC0 and TDC1 data choosing $3 < \Delta < 12$ would be acceptable.



\subsection{Cross-correlation}\label{subsec:cr-cor}
The cross-correlation between the light curves (two time series of data $X, Y$) in a system measures the \textit{correlation coefficient $\rho_{XY}$} in different time lags. This results in a set of correlation coefficients corresponding to different lags. The lag corresponding to the maximum coefficient gives the estimated time delay \citep{Peterson2001}. To derive these correlation coefficient at any time lag we need to have the data on both light curves corresponding to the assumed time lag and this cannot be the case since the data is not equi-spaced and it is sparse in cases. However, as discussed in subsection~\ref{subsec:smoothing} we tackle this issue by reconstructing the smooth forms of the light curves so we can measure the correlation coefficients at any given time lag. 

The Pearson correlation coefficient has the value between -1 to 1 and shows the linear dependency between the two random variables \citep{Wasserman2004}.
\begin{equation}\label{equ:cross}
	\rho_{XY}={\mathrm{cov}(X,Y) \over \sigma_X \sigma_Y}
\end{equation}
where $X$ and $Y$ indicate two time series with standard deviations $\sigma_X, \sigma_Y$ respectively and $\mathrm{cov}(X,Y)$ is their covariance.
The Pearson correlation coefficient can be estimated through the \textit{sample correlation coefficient, $r_{xy}$},
\begin{equation}\label{equ:sample_cross}
	r_{xy}=\frac{\sum\limits_{i=1}^n (x_i-\bar{x})(y_i-\bar{y})}
            {\sqrt{\sum\limits_{i=1}^n (x_i-\bar{x})^2 \sum\limits_{i=1}^n (y_i-\bar{y})^2}},
\end{equation}
where $x_i, y_i, (i=1,2,..,n)$ are data samples with sample means $\bar{x}$ and $\bar{y}$ of two time series $X, Y$ respectively \citep{Peterson2001}.
It can be also shown that the linear transformation of random variables $X, Y$ does not affect the correlation coefficient \citep{Wasserman2004}. Hence $\rho_{XY}=\rho_{X'Y'}$ if $X'=a_x X+b_x,$ $Y'=a_y Y+b_y$ where $a_x,$ $b_x,$ $a_y,$ $b_y$ are constant. 
This property has an important role in accurate estimation of the time delay when data suffers from microlensing, which we discuss in subsection~\ref{subsec:time_delay}.

\subsection{Time delay estimation}\label{subsec:time_delay}
\begin{figure}
   \includegraphics[width=0.9\textwidth]{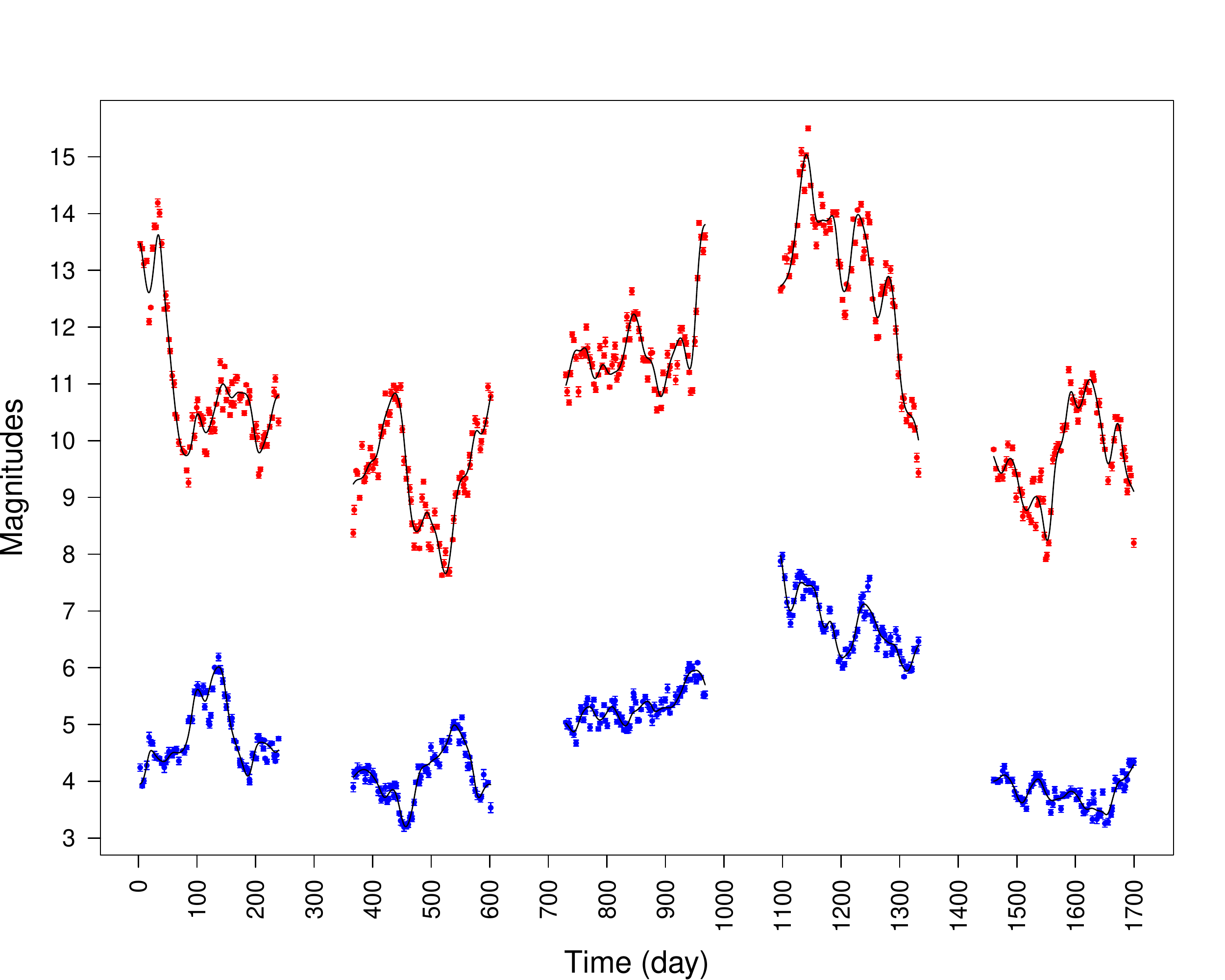}
  \caption{\label{fig:data_smooth} An example of TDC1 simulated light curves (rung 0, pair 2) with corresponding smooth fits (solid curves). We consider only the smooth fits in the regions that data is available.}
 \end{figure}

Given a pair of light curves $A_1, A_2$, we first apply the smoothing method (section~\ref{subsec:smoothing}) to obtain the smooth fits $A^{s}_1, A^{s}_2$. Figure~\ref{fig:data_smooth} depicts a pair of typical light curves and their smooth fits. We can see that the data are in segments and there are gaps between them. We should note that we consider only the data segments and their corresponding smooth fit. To calculate the cross-correlation we compare individual segments of the two light curves together
when there are at least three data points in the overlapped region. Next we sum up the correlation coefficients corresponding to each time lag and divide by the number of segments to obtain \textit{average correlation coefficient}.
 
To have a good control of the errors we calculate the cross-correlation between $A_1$ and $A^{s}_2$ as well as between $A_2$ and $A^{s}_1$. The best time delay in each case corresponds to the maximum correlation. This leads to two close values (absolute) for the time delay, $|\tilde{\Delta t}_{A_1;A^{s}_2}|$ and $|\tilde{\Delta t}_{A_2;A^{s}_1}|$. 
If the maximum average correlation coefficient in each case becomes more than $0.6$ we take their mean value, $|\tilde{\Delta t}_{A_{1}A_{2}}|$, and the standard deviation\footnote{We employ the \textit{sample standard deviation}, which is defined by
$\sigma = \sqrt{\frac{1}{n-1} \sum_{i=1}^n (x_i - \overline{x}_n)^2}\,$
where ${x_1,x_2,\ldots,x_n}$ is the sample and $\overline{x}_n$ is the sample mean. The equation~\ref{eq:initerr} is the same estimator for the case of $n=2$.}, $\sigma^{ini}_{A_{1}A_{2}}$, as the estimated time delay and its initial error. 

\begin{equation}
\label{eq:timedelay}
|\tilde{\Delta t}_{A_{1}A_{2}}|=\frac{|\tilde{\Delta t}_{A_1;A^{s}_2}|+|\tilde{\Delta t}_{A_2;A^{s}_1}|}{2}
\end{equation}

\begin{equation}
\label{eq:initerr}
\sigma^{ini}_{A_{1}A_{2}} =  \sqrt{(|\tilde{\Delta t}_{A_1;A^{s}_2}|-|\tilde{\Delta t}_{A_{1}A_{2}}|)^2 +  (|\tilde{\Delta t}_{A_2;A^{s}_1}|-|\tilde{\Delta t}_{A_{1}A_{2}}|)^2} = \sqrt{2} \times \frac{\left | |\tilde{\Delta t}_{A_1;A^{s}_2}|-|\tilde{\Delta t}_{A_2;A^{s}_1}| \right |}{2}
\end{equation}

\begin{figure}
   \includegraphics[width=\textwidth]{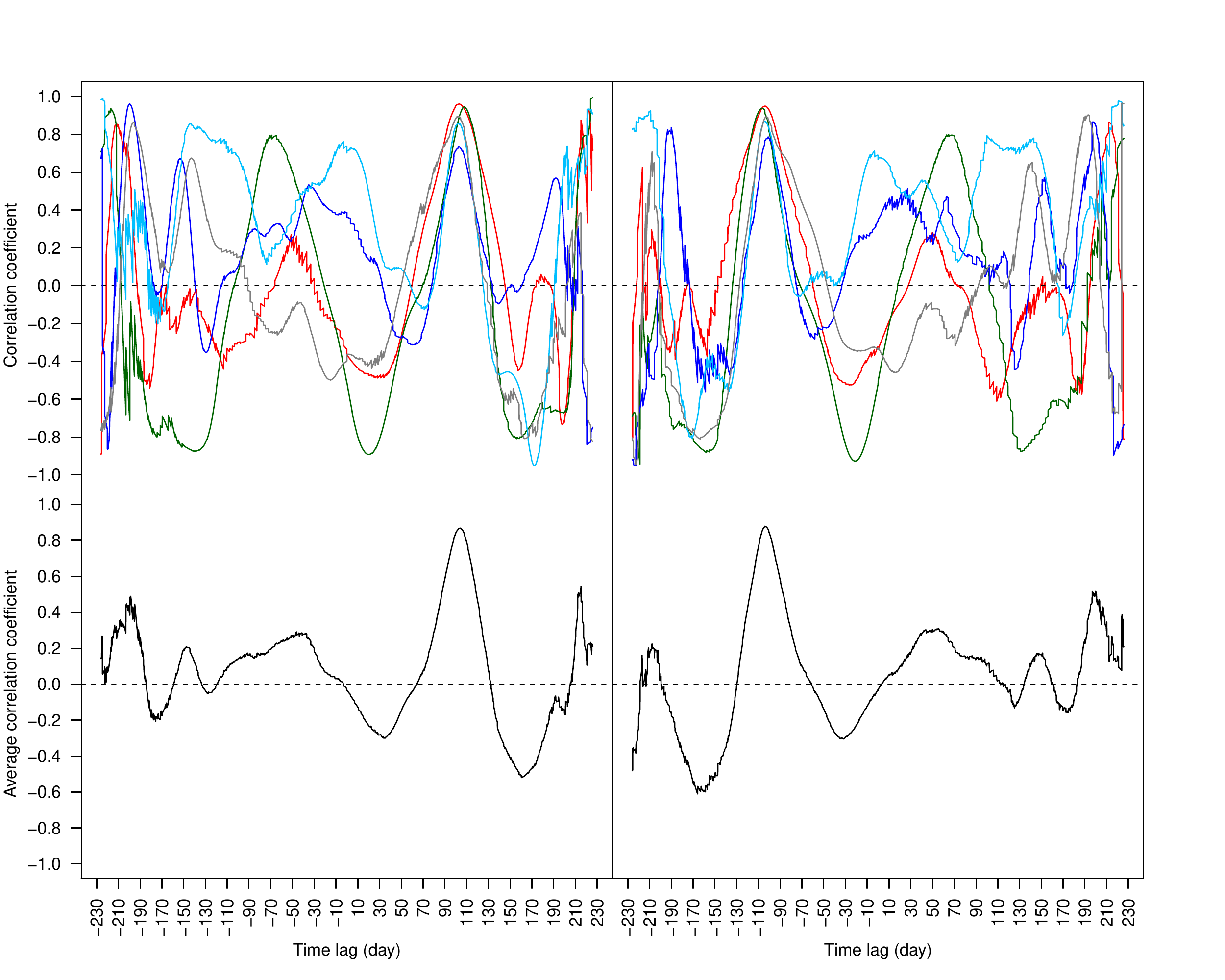}
  \caption{\label{fig:corr_lag} The upper-left and upper-right panels illustrate the cross-correlation of segments of the light curves data and smooth fits (illustrated in Figure \ref{fig:data_smooth}) which are obtained by comparing $A_1$ with $A^{s}_2$ and $A_2$ with $A^{s}_1$ respectively.
The lower panels depict corresponding average correlation coefficients. The corresponding best time delays are obtained $\tilde{\Delta t}_{A_1;A^{s}_2}= +103.15$ days and $\tilde{\Delta t}_{A_2;A^{s}_1}= -103.59$ days at maximum average correlation coefficients $0.87$ and $0.88$ respectively (lower panels).}
 \end{figure}

As an example, Figure \ref{fig:corr_lag} the upper-left and upper-right panels illustrate the cross-correlation of segments of the light curves data and smooth fits (illustrated in Figure \ref{fig:data_smooth}) obtained by comparing $A_1$ with $A^{s}_2$ and $A_2$ with $A^{s}_1$ respectively. 
The corresponding best time delays are obtained $\tilde{\Delta t}_{A_1;A^{s}_2}= +103.15$ days and $\tilde{\Delta t}_{A_2;A^{s}_1}= -103.59$ days at maximum average correlation coefficients $0.87$ and $0.88$ respectively (Figure \ref{fig:corr_lag}. lower panels). Using Equations~\ref{eq:timedelay} \& \ref{eq:initerr}, the time delay and its initial error are estimated as $|\tilde{\Delta t}_{A_{1}A_{2}}|=103.37$ days and $\sigma^{ini}_{A_{1}A_{2}}=0.31$ days respectively\footnote{The true time delay for this particular case is reported 104.22 days.}.

Our rejection criterion $\rho < 0.6$ is inferred from the feedbacks of our different submissions for simulated light curves in TDC0 challenge where we tried to achieve zero outlier and to avoid any single incorrect estimation of a time delay. 
We will elaborate on the error estimation in subsection~\ref{sec:error}.

Micolensing can generally cause distortions in light curves which creates difficulties in finding the time delay.
Since we compare the light curves in segments, the effect of microlensing can be considered as a linear distortion in different segments of the light curves. As we discussed in subsection~\ref{subsec:cr-cor} the correlation coefficient is not changed under the linear transformation of variables. Therefore there is no concern for microlensing in our methodology and this is one of the important advantages of our proposed method. Results of TDC0 and TDC1 challenge have shown that we have been correct in our argument as we could submit time delay estimates consistently for more than $30\%$ of the cases for all data rungs.

\subsection{Error estimation}\label{sec:error}
Apart from the derived initial uncertainty (subsection~\ref{subsec:time_delay}) of the estimated time delay for a pair of light curves (standard deviation of the two derived values of time delays comparing smooth fits and actual light curves) we need to consider other uncertainties. We should be very careful to avoid the effects of random fluctuations as they may mimic features or signals. Performing some extensive simulations, we found that even when $|\tilde{\Delta t}_{A_1,A_2^s}|$ is very close to $|\tilde{\Delta t}_{A_2,A_1^s}|$ we might still be considerably away from the actual time delay in some cases. This happens when the data is sparse with large gaps, and can result in a huge source of error in estimation of time delays. For each specific case one can make a number of simulations to understand the effect of the data distribution on the uncertainties of the time delay and this seems to be a proper approach. However since we had limited time for our analysis (we joined the challenge two months before its deadline), we chose a novel approach for our error estimation. We estimated the errors for different light curve pairs using an estimation of the errors using the rich information in the Quad systems. By using a consistency relation between various light curves of the Quad systems we calibrated our error estimation for different rungs with various time delays. In other words we can extract more information from Quad systems which can indirectly be used in our error estimation.

Suppose the four light curves of a Quad image are labeled $A_1, A_2, B_1$ and $B_2$. For every Quad system we should have
\begin{equation}
\label{eq:comb}
\tilde{\Delta t}_{A_{1}A_{2}} - (\tilde{\Delta t}_{A_{1}B_{1}} +\tilde{\Delta t}_{B_{1}A_{2}}) \pm \sqrt{(\sigma^{ini}_{\tilde{\Delta t}_{A_{1}A_{2}}})^2 + (\sigma^{ini}_{\tilde{\Delta t}_{A_{1}B_{1}}})^2 + (\sigma^{ini}_{\tilde{\Delta t}_{B_{1}A_{2}}})^2 } \equiv 0
\end{equation}
where $\tilde{\Delta t}_{A_{1}A_{2}}$ is the estimated time delay between the light curves $A_1$ and $A_2$ and $\sigma_{\tilde{\Delta t}_{A_1A_2}}$ is corresponding initial error (and same notation is used for other pairs). In short

\begin{equation}
\label{eq:difference}
T_{dif} \pm \sigma_{T_{dif}} \equiv 0
\end{equation}
where $T_{dif}=\tilde{\Delta t}_{A_{1}A_{2}} - (\tilde{\Delta t}_{A_{1}B_{1}} +\tilde{\Delta t}_{B_{1}A_{2}})$ and $\sigma_{T_{dif}}=\sqrt{(\sigma^{ini}_{\tilde{\Delta t}_{A_{1}A_{2}}})^2 + (\sigma^{ini}_{\tilde{\Delta t}_{A_{1}B_{1}}})^2 + (\sigma^{ini}_{\tilde{\Delta t}_{B_{1}A_{2}}})^2 }$. Now, if $\left|T_{dif}\right| \leqslant \sigma_{T_{dif}}$ then equation~\ref{eq:difference} is valid and therefore we can assume that all time delays and their corresponding errors are estimated consistently. If not, then the estimated errors are smaller than they should be. Hence we add an \textit{error correction} $\sigma_{ec}$ such that equation~\ref{eq:difference} will be satisfied.
The error correction would be calculated as the following
\begin{equation}
\sigma_{ec}^2 = \left|T_{dif}\right|^2 - \sigma_{T_{dif}}^2
\end{equation}
We add this error correction quadratically to the initial error. Therefore the new uncertainty for $\tilde{\Delta t}_{A_1A_2}$ would be
\begin{equation}
\sigma_{\tilde{\Delta t}_{A_{1}A_{2}}}^{new} = \sqrt{ (\sigma^{ini}_{\tilde{\Delta t}_{A_{1}A_{2}}})^2 + \frac{\alpha}{3}\sigma_{ec}^2}
\end{equation}
To be more conservative we choose $\alpha=2$ instead of $\alpha=1$ (which is an optimum consideration) in the above equation. 
The above argument and derivation is also valid for error estimation of $\tilde{\Delta t}_{B_{1}B_{2}}$.



\begin{figure}
   \includegraphics[width=0.9\textwidth]{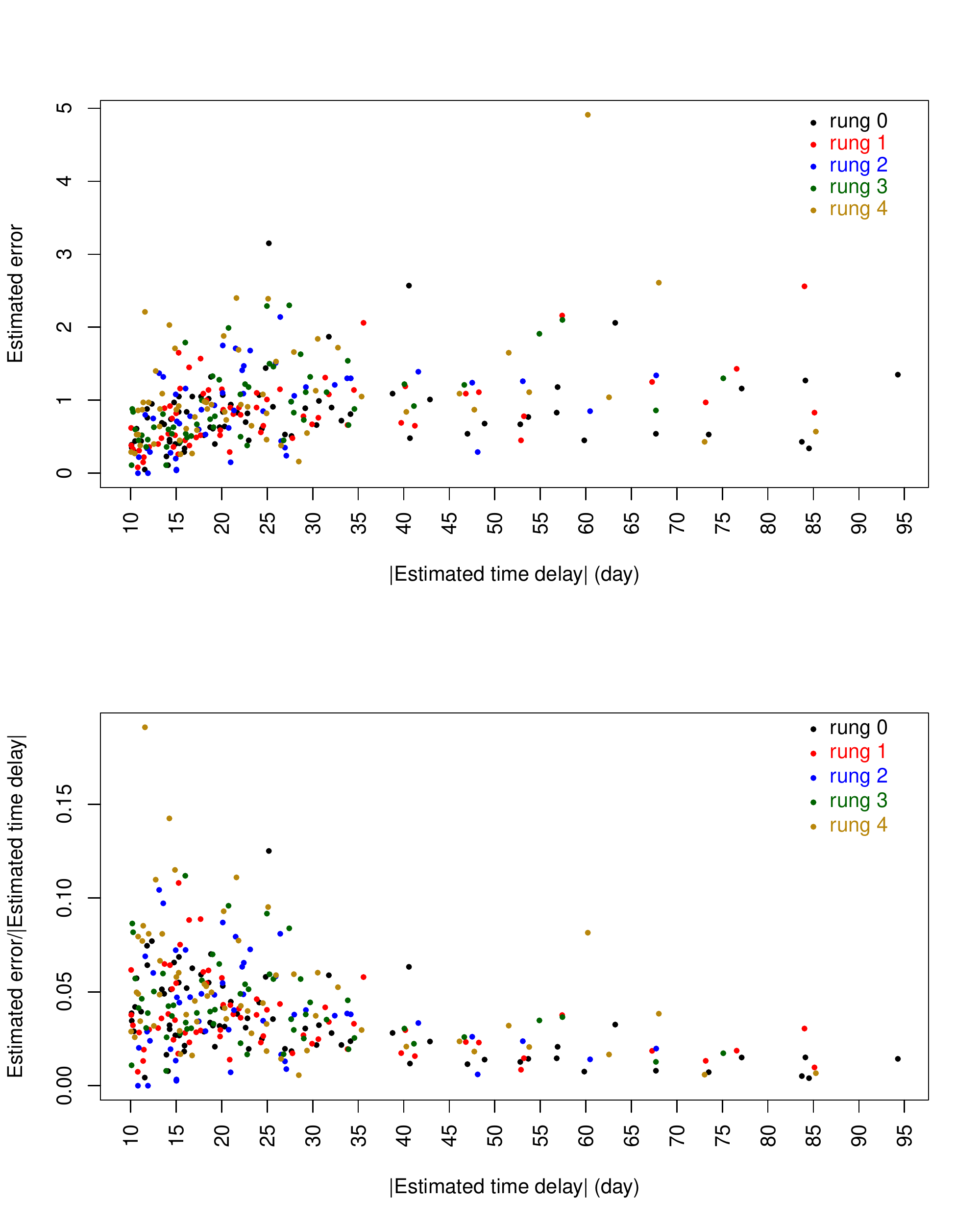}
  \caption{\label{fig:error_quad} The estimated errors and relative errors versus estimated time delays of Quad systems. These results are based on internal consistency of the estimated time delays in Quad systems and have been used to construct the error profile in subsection~\ref{sec:error}.}
 \end{figure}

Based on the above error estimation for Quad pairs we establish a profile to estimate the error for Double systems. We plot the estimated errors and relative errors versus estimated time delays (see Figure~\ref{fig:error_quad}) from the Quad systems for all rungs. This information suggests how large the uncertainties should be for different systems in different rungs depending on their estimated time delays. According to this information we suggest the uncertainties $\sigma_R$ for Double systems should be as the following:
\\*
For \textbf{Rung 0},\\

$
\text{if} \hspace{3mm} \lvert 	\tilde{\Delta t} \lvert \leqslant 20 \Rightarrow
\sigma_R=0.06 \times \lvert \tilde{\Delta t} \lvert
$
, \hspace{15mm}
$
\text{if} \hspace{3mm} \lvert \tilde{\Delta t} \lvert > 20 \Rightarrow
\sigma_R=1.2
$
\vspace{10mm}
\\*
For \textbf{Rung 1},\\

$
\text{if} \hspace{3mm} \lvert \tilde{\Delta t} \lvert \leqslant 20 \Rightarrow
\sigma_R=0.06 \times \lvert \tilde{\Delta t} \lvert
$
, \hspace{15mm}
$
\text{if} \hspace{3mm} \lvert \tilde{\Delta t} \lvert > 20 \Rightarrow
\sigma_R=1.3
$
\vspace{10mm}
\\*
For \textbf{Rung 2},\\

$
\text{if} \hspace{3mm} \lvert \tilde{\Delta t} \lvert \leqslant 30 \Rightarrow
\sigma_R=0.07 \times \lvert \tilde{\Delta t} \lvert
$
, \hspace{15mm}
$
\text{if} \hspace{3mm} \lvert \tilde{\Delta t} \lvert > 30 \Rightarrow
\sigma_R=1.3
$
\vspace{10mm}
\\*
For \textbf{Rung 3},\\

$
\text{if} \hspace{3mm} \lvert \tilde{\Delta t} \lvert \leqslant 30 \Rightarrow
\sigma_R=0.08 \times \lvert \tilde{\Delta t} \lvert
$
, \hspace{15mm}
$
\text{if} \hspace{3mm} \lvert \tilde{\Delta t} \lvert > 30 \Rightarrow
\sigma_R=1.5
$
\vspace{10mm}
\\*
For \textbf{Rung 4},\\

$
\text{if} \hspace{3mm} \lvert \tilde{\Delta t} \lvert \leqslant 25 \Rightarrow
\sigma_R=0.08 \times \lvert \tilde{\Delta t} \lvert
$
, \hspace{15mm}
$
\text{if} \hspace{3mm} \lvert \tilde{\Delta t} \lvert > 25 \Rightarrow
\sigma_R=1.5
$
\vspace{5mm}
\\*
One should note that we used these simple boundaries for simplification and in principle one can directly look at the profile of the results from Quad systems to estimate the expected uncertainties for a Double system with a particular time delay in a specific rung. To be conservative, these estimated errors have been added to the initial errors quadratically. So finally, the errors we considered for our submissions in the TDC1 challenge have been (for any pair $A_{1}A_{2}$):
\begin{equation}
\label{eq:finalerror}
\sigma_{\tilde{\Delta t}_{A_{1}A_{2}}} = \sqrt{ (\sigma^{ini}_{\tilde{\Delta t}_{A_{1}A_{2}}})^2 + \sigma_{R}^2 }
\end{equation}

\section{Results and discussion}\label{sec:results_discussion}
To evaluate the results of the challenge Evil team used the following metrics \citep{TDC1-2014}:\\
$f$ is the fraction of the submitted time delays,
\begin{equation}
f\equiv \frac{N_{submitted}}{N}
\end{equation}
$\chi^2$ shows the average weighted distance between the estimated and true time delays,
\begin{equation}
\chi^2=\frac{1}{fN}\sum_{i}\left ( \frac{ \tilde{\Delta t_i}-\Delta t_i}{\sigma_i} \right )^2
\end{equation}
$P$ is the average precision metric,
\begin{equation}
P=\frac{1}{fN}\sum_{i}\left ( \frac{\sigma_i}{|\Delta t_i|} \right )
\end{equation}
and $A$ reflects the average bias of estimation,
\begin{equation}
A=\frac{1}{fN}\sum_{i}\left ( \frac{ \tilde{|\Delta t_i|}-|\Delta t_i|}{|\Delta t_i|} \right )
\end{equation}
where $N$ is the total number of the light curves for analysis, $N_{submitted}$ is the number of submitted time delays, $\Delta t_i$ is true time delay and $\tilde{\Delta t_i}$ is estimated time delay with uncertainty $\sigma_i$. The criteria for passing the TDC0 stage were $f>0.3$, $P<0.15$, $A<0.15$ and $\chi^2<2$. We passed TDC0 stage in our first attempt and working with TDC1 data, since there was no fixed criteria, our rational expectation was that we should set the TDC0 limits of the metrics as the lower boundaries of the TDC1 criteria. Hence we focused to improve our results in all four criteria.      

Table~\ref{tab:feedback} shows the feedback of our results given by TDC1 challenge Evil team. For all rungs we achieved $f>0.3$, which was one the main criteria of passing the TDC0 stage of the challenge. This table clearly shows that despite of having a considerably large $f$ factor, our results do not include numerous outliers and our algorithm estimates the time delays reliably and precisely. It is worth emphasizing that all steps of our algorithm are fully automated and there is no need for direct human evaluation at any stage. Results of this table shows that we could achieve very good $f$ factor with $f>0.3$ for all rungs and having precise measurements of time delays and their uncertainties with $\chi^2<1$ and $P<0.06$ for all rungs. However, we can see that there is a slight bias in our results, as reflected in the $A$ metric. 

\begin{figure}
   \includegraphics[width=0.9\textwidth]{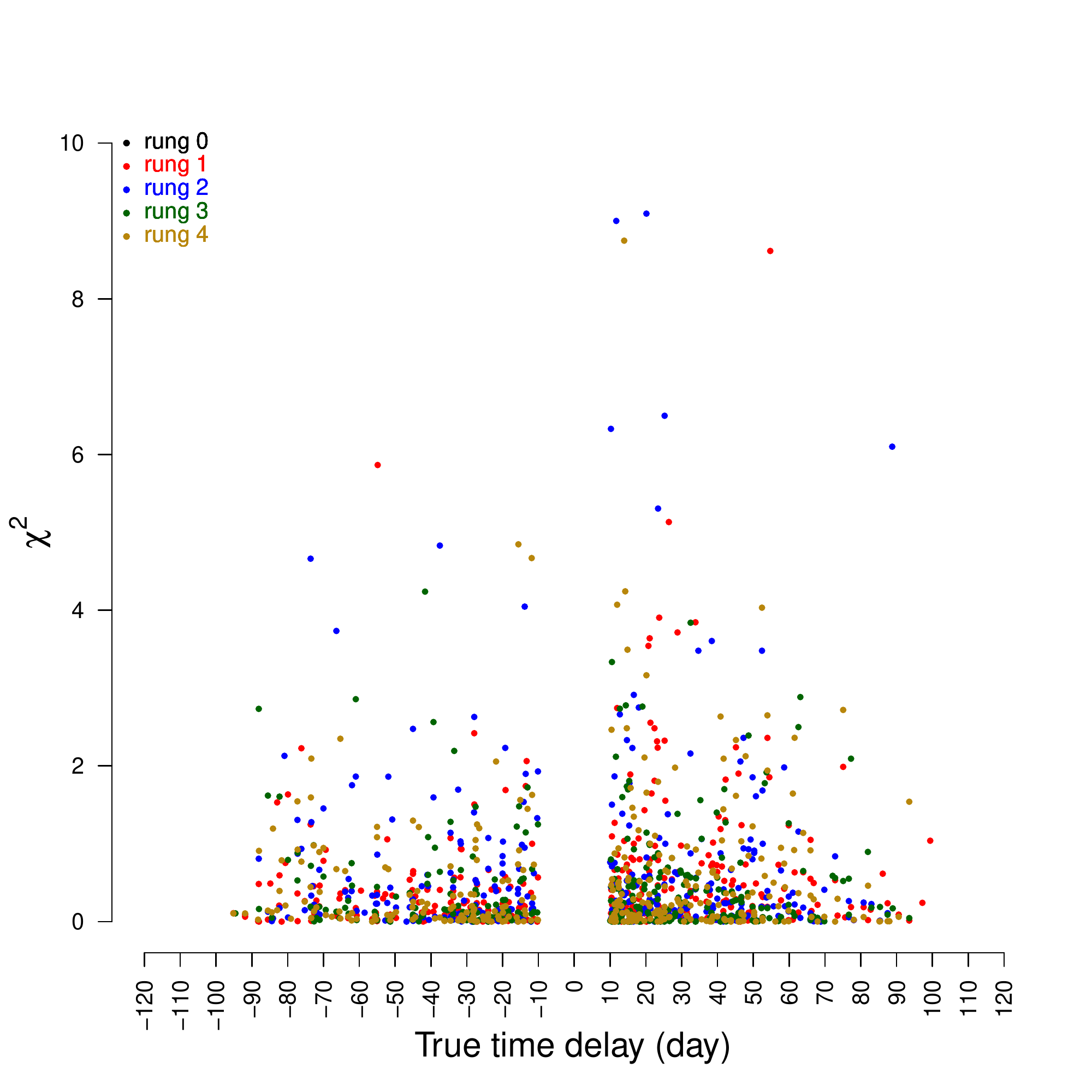}
  \caption{\label{fig:chi2_vs_true} The $\chi^2$ values for all entries with $\chi^2<10$ (the entries of rung 0, 1, 2, 3, 4 are colored black, red, blue, green and gold respectively). Within all rungs there have been only 5 entries with $\chi^2>10$ but we consider this boundary to be consistent with TDC1 publication. }
 \end{figure}

\begin{figure}
   \includegraphics[width=0.9\textwidth]{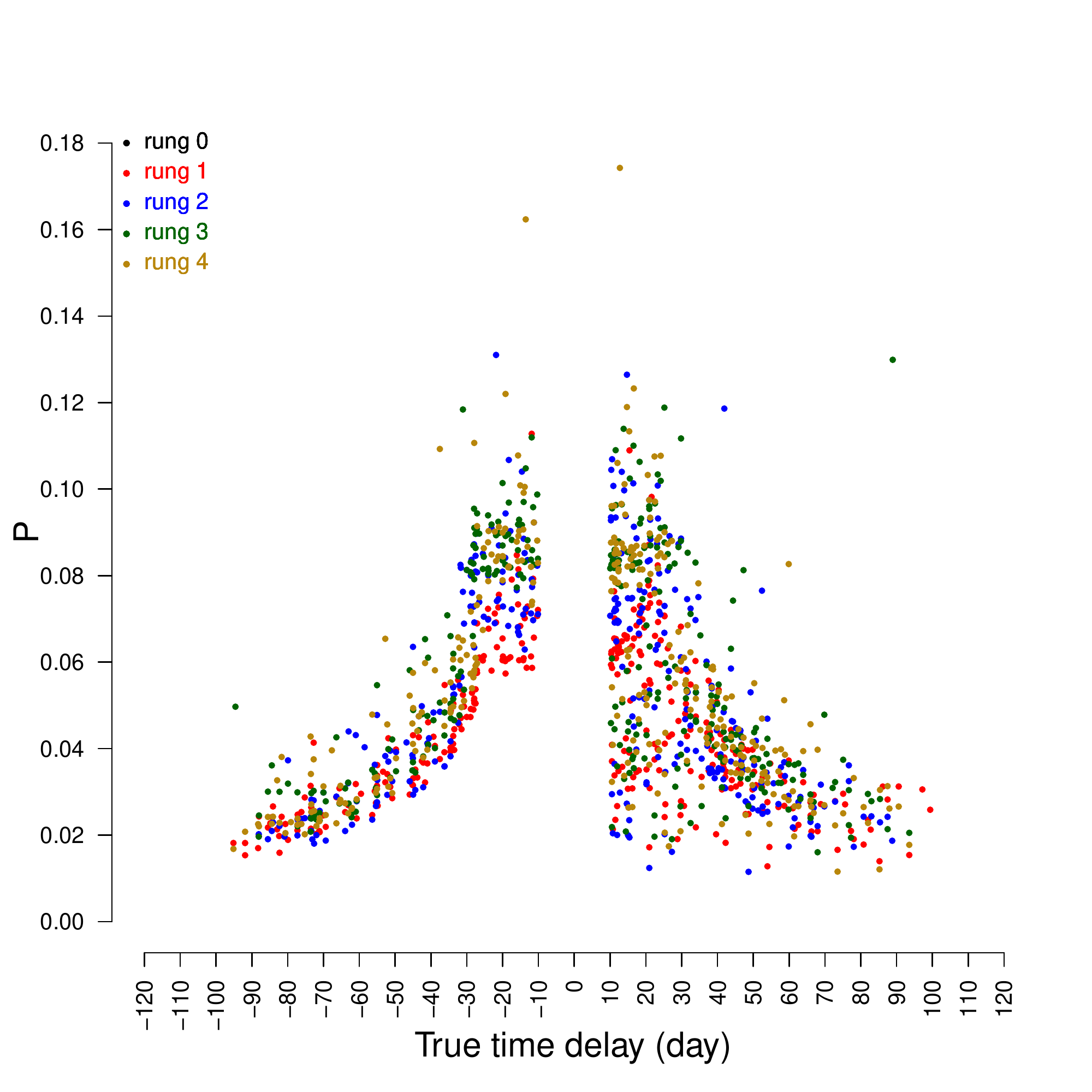}
  \caption{\label{fig:P_vs_true} The P values for all entries with $\chi^2<10$ (the entries of rung 0, 1, 2, 3, 4 are colored black, red, blue, green and gold respectively). Within all rungs there have been only 5 entries with $\chi^2>10$ but we consider this boundary to be consistent with TDC1 publication.}
 \end{figure}

\begin{figure}
   \includegraphics[width=0.9\textwidth]{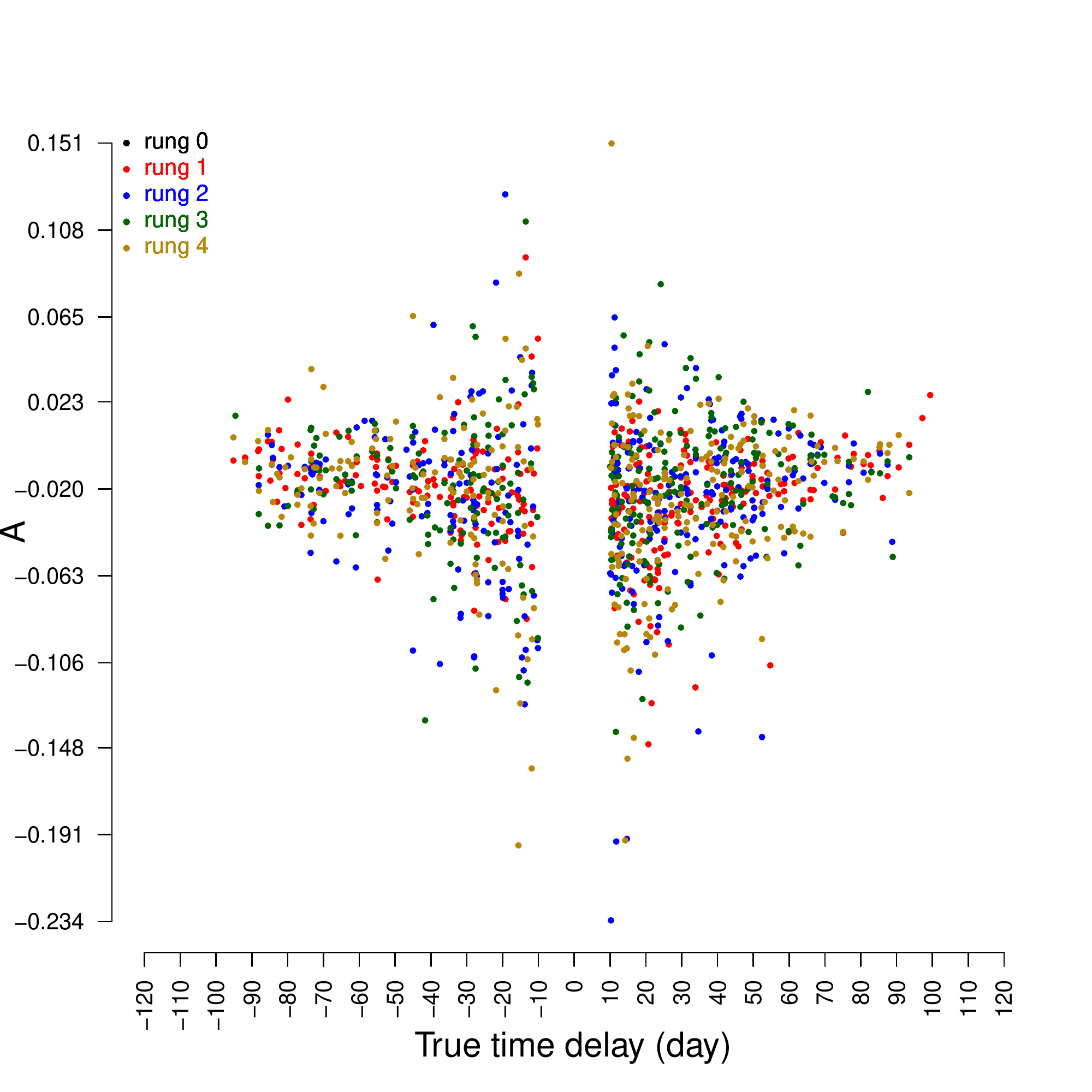}
  \caption{\label{fig:A_vs_true} The A values for all entries with $\chi^2<10$ (the entries of rung 0, 1, 2, 3, 4 are colored black, red, blue, green and gold respectively). Within all rungs there have been only 5 entries with $\chi^2>10$ but we consider this boundary to be consistent with TDC1 publication.}
 \end{figure}

\begin{figure}
   \includegraphics[width=\textwidth]{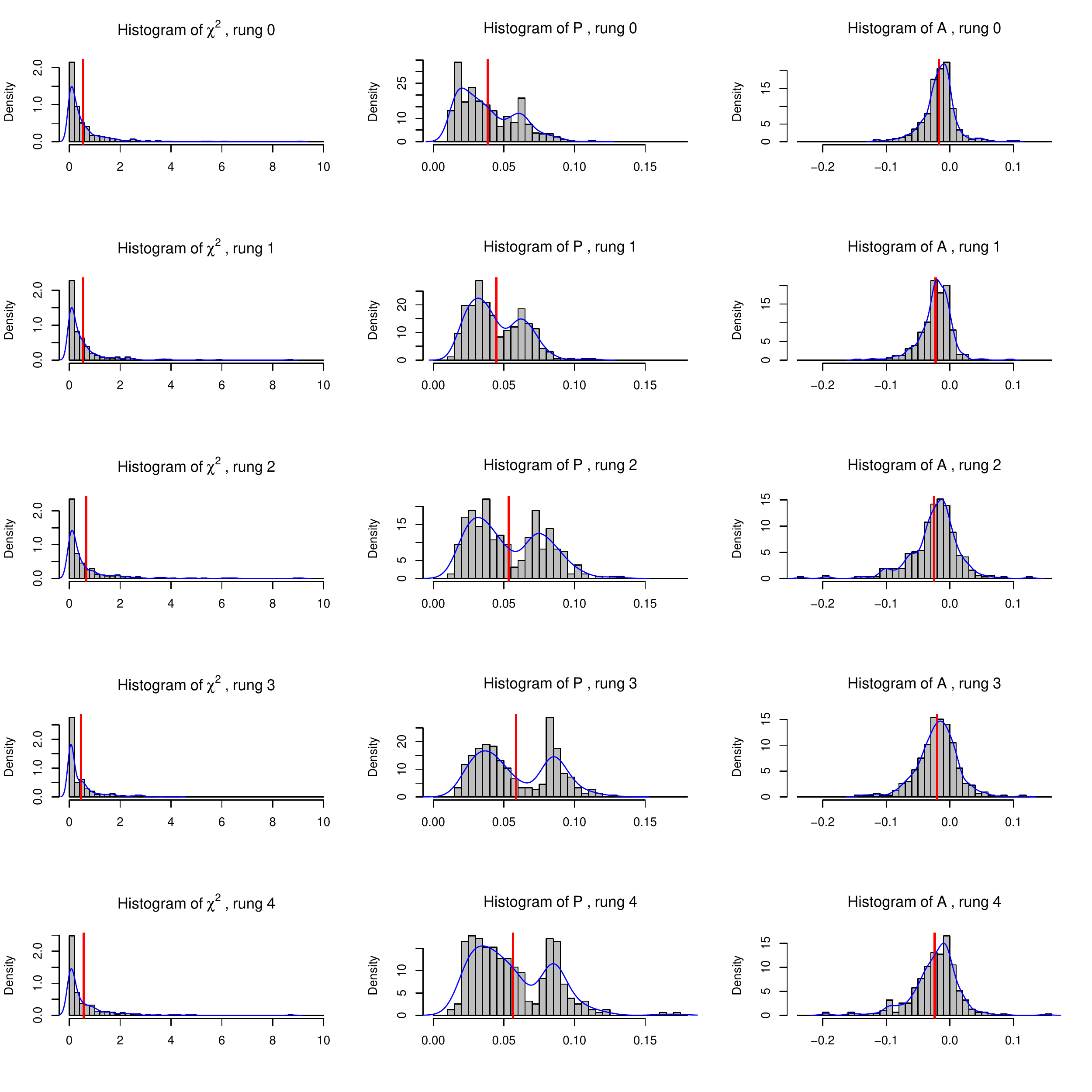}
  \caption{\label{fig:hist_rung0} The histograms of $\chi^2$, P and A for every rung separately. The blue curve shows the corresponding kernel densities. The mean values are indicated with vertical red line.}
 \end{figure}

\begin{figure}
   \includegraphics[width=1.1\textwidth]{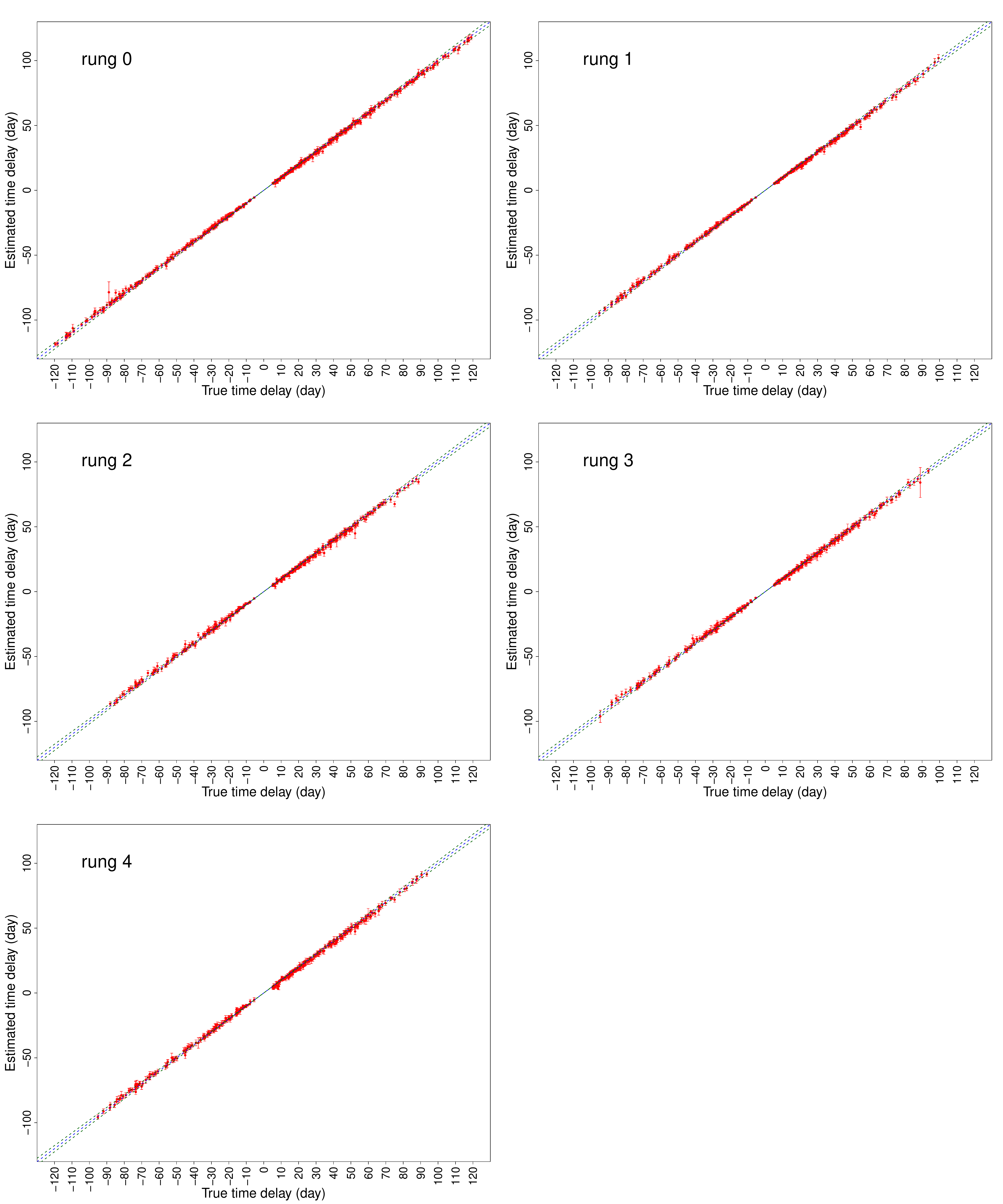}
  \caption{\label{fig:fitted_true} The estimated versus true time delays for {\it all} submitted entries {\it without} the $\chi^2<10$ cut-off. One can see clearly that method introduce no outlier in neither of the rungs while keeping $f>0.3$ in all cases.}
\end{figure}

The Figures~\ref{fig:chi2_vs_true}, \ref{fig:P_vs_true}, \ref{fig:A_vs_true} depict the $\chi^2, P, A$ individually for all the entries which have $\chi^2<10$ (the entries of rung 0, 1, 2, 3, 4 are colored black, red, blue, green and gold respectively). We choose the $\chi^2<10$ cut-off to be consistent with Evil team settings in TDC1 paper \citep{TDC1-2014}. Our results show that there are only 5 items with $\chi^2\ge10$ out of few thousand entries (within all rungs) which is indeed a very small fraction and statistically insignificant. We have plotted the histograms of the metrics in Figure~\ref{fig:hist_rung0}. In Figure~\ref{fig:fitted_true} the estimated time delays versus true time delays are plotted. For comparison $45^{\circ}$ line and its $2\%$ deviations are also plotted. This plot clearly shows the consistency and power of our method where there is no outlier in any of the rungs and in almost all cases we had a reliable and precise estimation of time delays. 

As mentioned earlier our results had a small bias ($A$ was derived to be approximately $0.018$ at all rungs) which indicates a need for further calibration. However, after the true time delay values of TDC0 and TDC1 challenges were revealed we noticed that our time delay estimations in all rungs had a constant $0.5$ day bias. This constant bias appeared due to a simple fact that the TDC0 rung 0 simulated data was equi-spaced with 1 day gaps between neighboring data points and we calibrated the method to find time delays on these integer dates. So in fact the bias in our results is due to a calibration issue rather than being a bias in the method. Hence by simply adding 0.5 days to all estimated time delays $A$ was improved to $0.1-0.6\%$, which is very much acceptable. 

On the other hand, we noticed that our average $\chi^2$ is very small at all rungs which indicates that we could have a smaller $P$ metric if we had smaller errors, while $\chi^2$ could be enlarged slightly but still within an acceptable range. Hence by dividing all estimated errors by a factor of $\sqrt{2}$ the average $P$ value improved to $0.027 \sim 0.041$ while the average $\chi^2$ reached to around $1$. The calibrated results are tabulated in Table~\ref{tab:calibrated}. These results show that with a minor modification the algorithm could yield a reliable and precise time delay estimation for various forms of the data and in a fully automated manner. 

\section{Conclusion}\label{section:conclusion}
In this paper we present the algorithm that we used to participate in TDC0 and TDC1 Strong Lens Time Delay Challenges \citep{TDC0-2013, TDC1-2014}. In our algorithm we combined various statistical methods of data analysis in order to estimate the time delay between different light curves of strong lens systems. At different stages of our analysis we used iterative smoothing, cross-correlation, simulations and error estimation, bias control and significance testing to prepare our results. Given the limited timeframe, we had to make some approximations in our error analysis. However the novel approach we used to estimate the uncertainties has been shown to perform satisfactorily.
In our approach to estimate the time delay between a pair of light curves $A_1$ and $A_2$, we first smoothed over both light curves using an iterative smoothing method \citep{Arman2006,Arman2007,Arman2010,Arman2012}, producing the smoothed light curves $A^{s}_1$ and $A^{s}_2$. During smoothing, we recorded the ranges with no data available (which would have resulted in unreliable smoothing). The algorithm
was set to detect such ranges automatically. Then, we calculated the cross-correlation between $A_1$ and $A^{s}_2$ and also between $A_2$ and $A^{s}_1$ for different time delays, and found the maximum correlations. These two maximum correlations should be for the same time delays (that is, the absolute values of the time delays should be consistent with each other). The difference between these two estimated time delays (with maximum correlations) was part of the total uncertainty considered for each pair (in the estimated time delay). To estimate the error on each derived time delay, we simulated many realizations of the data for each rung, and for various time delays. Knowing the fiducial values, we derived the expected uncertainties in the estimated values of the time delays. In the case of the Quad sample, we used different combinations of the smoothed and raw light curves to test the internal consistency of the results and the relative errors. These internal consistency relations were used to adjust the estimated error-bars for each pair. In our final submission we used these internal consistencies between the estimate time delays within Quad systems for our error estimations.  For the bias control, since long time delays have a limited data overlap between the two light curves we only passed systems with cross-correlations with a correlation coefficient getter than $0.6$. Throughout this challenge one of our main concerns has been to achieve a high $f$ value without any outliers. This was achieved with $f > 0.3$ for all five rungs. Our conservative approach yielded average $\chi^2$ values of around $0.5$ to $0.9$ for different rungs with $P$ of approximately $0.03$ to $0.06$. Since $\chi^2$ and $P$ are correlated, by simply dividing all estimated errors by a factor of $\sqrt 2$, $\chi^2$ of $\sim1$ and $P$ of $0.02$ to $0.04$ could be achieved trivially. Since our method was initially calibrated with only on TDC0 rung 0 data, owing to lack of time, a calibration bias of $0.5$ days for all the submissions was discovered, resulting in $A \approx 1.8-2.5\%$. By adding this calibration correction of $0.5$ days to all our submissions' time delay estimations, the bias was removed, improving $A$ substantially to only $0.1-0.6\%$. It has been discussed recently~\citep{Hojjati2014} that it is very important to have a good bias control in time delay estimation in order to have a reasonable cosmology determination. To summarize, our proposed method seems promising in both reliability and precision, and is automated in all steps. It is also fast with processing time in the order of few minutes (using a typical PC) for a pair of light curves.
The method presented here is exactly the same algorithm we used to participate in TDC challenges, and there are ways to improve our error estimation by doing appropriate simulations for each set of light curves separately. Significant modifications of this algorithm will be reported in future publications.

\section{Acknowledgments}
The authors would like to thank Eric Linder for various discussions throughout this work. Authors would like to thank Shi Won Ahn for computational support and Alireza Hojjati, Stephen Appleby, Hyungsuk Tak and Tommaso Treu for their comments on our manuscript. We also thank the organizers of the Strong Lens Time Delay Challenge ``Evil Team'' (K. Liao, G. Dobler, C. D. Fassnacht, P. Marshall, N. Rumbaugh, T. Treu) for providing the simulated light curves (available at http://timedelaychallenge.org) and feedback during the challenges. We would also thank both Evil and Good teams of the Time Delay Challenge for useful discussions. A.A and A.S wish to acknowledge support from the Korea Ministry of Education, Science and Technology, Gyeongsangbuk-Do and Pohang City for Independent Junior Research Groups at the Asia Pacific Center for Theoretical Physics. A.S. would like to acknowledge the support of the National Research Foundation of Korea (NRF-2013R1A1A2013795).

\begin{table}[!htb]
\begin{center}
\vspace{6pt}
\resizebox{8cm}{!} {
\begin{tabular}{| c | c | c | c | c |}
\hline
Rung & f & $\chi^2$  & P & A  \\
\hline
0 & 0.529 & 0.579 & 0.038 & -0.018 \\ 
1 & 0.366 & 0.543 & 0.044 & -0.022 \\ 
2 & 0.350 & 0.885 & 0.053 & -0.025  \\ 
3 & 0.337 & 0.524 & 0.059 & -0.021  \\ 
4 & 0.346 & 0.608 & 0.056 & -0.024 \\
\hline
\end{tabular}
}
\end{center}\caption{~\label{tab:feedback} The feedback of our results (blind submission) given by TDC1 challenge Evil team. The $f > 0.3$, with $\chi^2<1$ and $P<0.06$ reflect a precise measurements of time delays and their uncertainties for all rungs of the data. There is a slight bias as it is emerged in the A metric.}
\end{table}

\begin{table}[!htb]
\begin{center}
\vspace{6pt}
\resizebox{8cm}{!} {
\begin{tabular}{| c | c | c | c | c |}
\hline
Rung & f & $\chi^2$  & P & A  \\
\hline
0 & 0.529 & 0.792 & 0.027 & -0.0014\\
1 & 0.366 & 0.660 & 0.031 & -0.0036\\
2 & 0.350 & 1.439 & 0.038 & -0.0058\\
3 & 0.337 & 0.766 & 0.041 & -0.0010\\
4 & 0.346 & 0.868 & 0.040 & -0.0048\\
\hline
\end{tabular}
}
\end{center}\caption{~\label{tab:calibrated} The overall calibrated results. 0.5 day is added to \textit{all} time delay estimations in order to remove the bias (explained in the text) and \textit{all} estimated errors are divided by a factor of $\sqrt{2}$. These calibrations result to a dramatic improvement for all these metrics.}
\end{table}

\bibliographystyle{apj}
\bibliography{../ref}

\end{document}